# Leveraging Large Language Models to Enhance Personalized Recommendations in E-commerce

1st Wei Xu †
*Independent Researcher*
Los Altos, USA
williamxw09@gmail.com

2nd Jue Xiao †
*Independent Researcher*
Jersey City, USA
juexiaowork@gmail.com

3rd Jianlong Chen* †
*Independent Researcher*
Beijing, China
jianlong.chen@ieee.org

***Abstract*—This study deeply explores the application of large language model (LLM) in personalized recommendation system of e-commerce. Aiming at the limitations of traditional recommendation algorithms in processing large-scale and multi-dimensional data, a recommendation system framework based on LLM is proposed. Through comparative experiments, the recommendation model based on LLM shows significant improvement in multiple key indicators such as precision, recall, F1 score, average click-through rate (CTR) and recommendation diversity. Specifically, the precision of the LLM model is improved from 0.75 to 0.82, the recall rate is increased from 0.68 to 0.77, the F1 score is increased from 0.71 to 0.79, the CTR is increased from 0.56 to 0.63, and the recommendation diversity is increased by 41.2%, from 0.34 to 0.48. LLM effectively captures the implicit needs of users through deep semantic understanding of user comments and product description data, and combines contextual data for dynamic recommendation to generate more accurate and diverse results. The study shows that LLM has significant advantages in the field of personalized recommendation, can improve user experience and promote platform sales growth, and provides strong theoretical and practical support for personalized recommendation technology in e-commerce.**

## I. Introduction

### A. Research background and importance

The rapid development of e-commerce has made personalized recommendation systems an important means to improve user experience and increase sales. However, traditional recommendation algorithms have certain limitations when dealing with large-scale, multi-dimensional data, and it is difficult to accurately capture users' deep needs and interests, resulting in insufficient accuracy and diversity of recommendation effects [1]. The capabilities of advanced graph attention networks in efficiently classifying complex data structures within heterogeneous information networks [2] highlight the potential of large language models to effectively process multidimensional data in e-commerce. In recent years, with the rise of large language models (LLMs), the field of natural language processing has made significant progress. LLMs have begun to attract widespread attention from academia and industry due to their outstanding performance in language understanding and generation, and have gradually been applied to personalized recommendation systems [3].

LLMs have powerful semantic understanding and generation capabilities. They can reveal users' potential interests and needs by processing and analyzing a large amount of natural language text generated by users, thereby significantly improving the performance of recommendation systems. For example, by using LLMs to conduct in-depth analysis of user text input, users' multi-level needs can be better captured and more accurate recommendation results can be generated [4]. In addition, LLMs can effectively utilize external knowledge to make up for the shortcomings of traditional recommendation systems in data sparsity and cold start problems, thereby providing more personalized and flexible recommendation services [5]. Using LLMs to improve the effectiveness of personalized recommendation systems not only helps to improve the accuracy and diversity of recommendations, but also provides a new technical approach to address the challenges faced by traditional recommendation algorithms. With the continuous deepening of the application of LLMs in recommendation systems, their importance and potential value are becoming more and more significant [6].

### B. Research Objectives

This study aims to explore how to use large language models (LLMs) to improve the effectiveness of personalized recommendations in e-commerce. First, this paper analyzes the current development status of existing personalized recommendation technologies and LLMs through a systematic literature review, and discusses in detail the advantages and disadvantages of existing methods. Secondly, this paper designs a recommendation system architecture based on LLMs, focusing on how to use the natural language processing capabilities of LLMs to improve the performance of recommendation systems in capturing users' multi-level needs. Finally, this paper verifies the designed recommendation system architecture through data collection, model training and experimental analysis, aiming to explore its effectiveness in improving recommendation accuracy and diversity. The ultimate goal of this study is to provide a new methodological support for personalized recommendation systems in e-commerce and to provide a useful reference for further exploration of LLMs in practical applications.

## II. Literature Review

### A. Current Status and Development of Personalized Recommendation Technology in E-commerce

Personalized recommendation technology in e-commerce has made significant progress in recent years. With the rapid growth of the number of goods and users on e-commerce platforms, traditional recommendation systems, such as collaborative filtering, content filtering, and rule-based recommendation, have gradually faced performance bottlenecks, especially in terms of data sparsity and recommendation accuracy [7]. In order to meet these challenges, researchers have begun to explore more complex and intelligent recommendation algorithms, such as recommendation systems based on deep learning and big data analysis. These new recommendation systems can better handle large-scale data and perform well in terms of recommendation accuracy and real-time performance [8]. The introduction of machine learning technology has also greatly enhanced the capabilities of personalized recommendation systems. By leveraging users' historical behavior data and product attribute information, these systems can generate more accurate recommendation results. For example, hybrid recommendation systems that combine offline data mining and online real-time recommendation are widely used in modern e-commerce platforms to improve the system's response speed and recommendation effect [9]. In addition, with the development of big data technology, personalized recommendation systems can provide more customized recommendation services by deeply analyzing user behavior [10]. However, despite many advances, existing personalized recommendation systems still face some challenges, such as data privacy issues, insufficient diversity of recommendation results, and algorithm interpret ability. Solving these problems will be an important direction for future research and the key to improving e-commerce personalized recommendation technology [11].

† These authors contributed equally to this work.

### B. *Development and Application of Large Language Models (LLMs)*

Large Language Models (LLMs) have made significant progress in the field of natural language processing in recent years, especially in language understanding and generation tasks. LLMs can learn rich semantic information by pre-training on large-scale text data and show strong adaptability in a variety of downstream tasks [12]. The scale and complexity of these models continue to increase with the improvement of computing power, so that they can handle more complex language tasks. At the application level, LLMs have been widely used in many fields such as dialogue systems, text generation, and machine translation. For example, models such as ChatGPT perform well in tasks such as generating natural dialogues, writing code, and providing real-time translation services, significantly improving the naturalness and efficiency of user interactions. In addition, LLMs have found applications in areas such as financial research, transportation flow prediction, healthcare, and creative tasks like lyric generation, further showcasing their versatility across different domains [13] [14] [15] [16]. LLMs are also used in the field of education to help students improve their learning outcomes by automatically generating learning content and providing personalized teaching suggestions [17]. However, the development of LLM is also accompanied by some challenges, such as huge consumption of computing resources, data bias and ethical issues. Solving these problems requires further research and technological innovation to ensure that the widespread application of LLM will not bring negative effects [18].

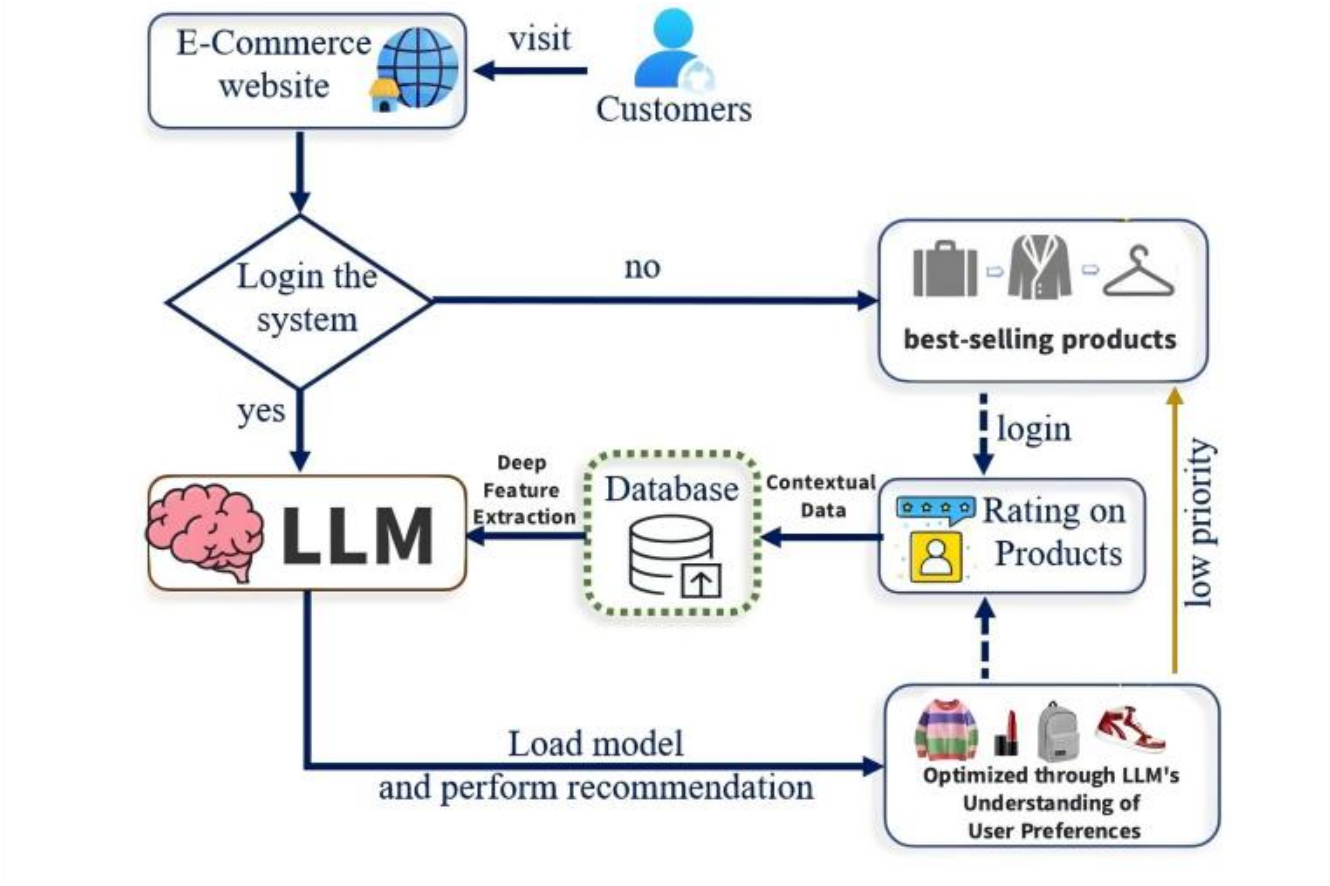

Fig. 1. Application of LLM in e-commerce platform recommendation system

Fig. 1 shows the workflow of the recommendation system of an e-commerce platform [19]. When a customer visits a website, the system first checks whether the user is logged in. For logged-in users, the system calls a personalized recommendation model based on a large language model (LLM), filters relevant data from the database, loads the recommendation model, and generates personalized recommended products. This process utilizes the deep semantic understanding capabilities of LLM to provide more accurate recommendations by analyzing the user's historical behavior and contextual information. For users who are not logged in, the system displays all products or popular products, and allows users to evaluate products after logging in. These evaluations will be saved and used to optimize future recommendation results.

### C. *Current research status of LLM application in personalized recommendation*

As the application of large language models in natural language processing gradually matures, researchers have begun to explore its potential in personalized recommendation systems. LLM's advantages in understanding and generating natural language text make its application in personalized recommendation possible. By combining user-generated text data with product attribute data, LLM can better capture user interests and needs, thereby improving the accuracy and diversity of recommendations [3]. At present, some studies have attempted to combine LLM with traditional recommendation algorithms to form a hybrid recommendation system to enhance the recommendation effect. For example, some studies use LLM to generate user portraits to more accurately identify users' long-term interests and short-term needs, which shows obvious advantages in cold start problems [20]. In addition, LLM's ability to process multimodal data (such as text, images, videos, etc.) enables it to play a wider role in recommendation systems, such as generating personalized content recommendations or interactive recommendation experiences [21]. Although LLM has broad application prospects in personalized recommendation, there are still many problems to be solved, such as how to effectively integrate LLM with existing recommendation algorithms and how to reduce the impact of data bias on recommendation results. Solving these problems will help improve the overall performance of personalized recommendation systems and further promote the application of LLM in recommendation systems [3].

## III. Research Design

### A. Research Framework Design

The research framework includes three core modules (as shown in Fig. 2): data layer, model layer and application layer. These modules are closely connected through data flow and model integration to jointly build an efficient personalized recommendation system.

The whole framework is described by the following formula:

$$R_{u,i} = \alpha \cdot f_{trad}(u,i) + \beta \cdot f_{LLM}(u,i) + \gamma \cdot f_{hybrid}(u,i) \tag{1}$$

Where $R_{u,i}$ represents the recommendation score of user $u$ for product $i$, $f_{trad}(u,i)$ is the score calculated by the traditional recommendation algorithm, $f_{LLM}(u,i)$ is the score of the LLM model, and $f_{hybrid}(u,i)$ is the combined score of the two. The parameters $\alpha$, $\beta$, and $\gamma$ represent the weights of each part respectively.

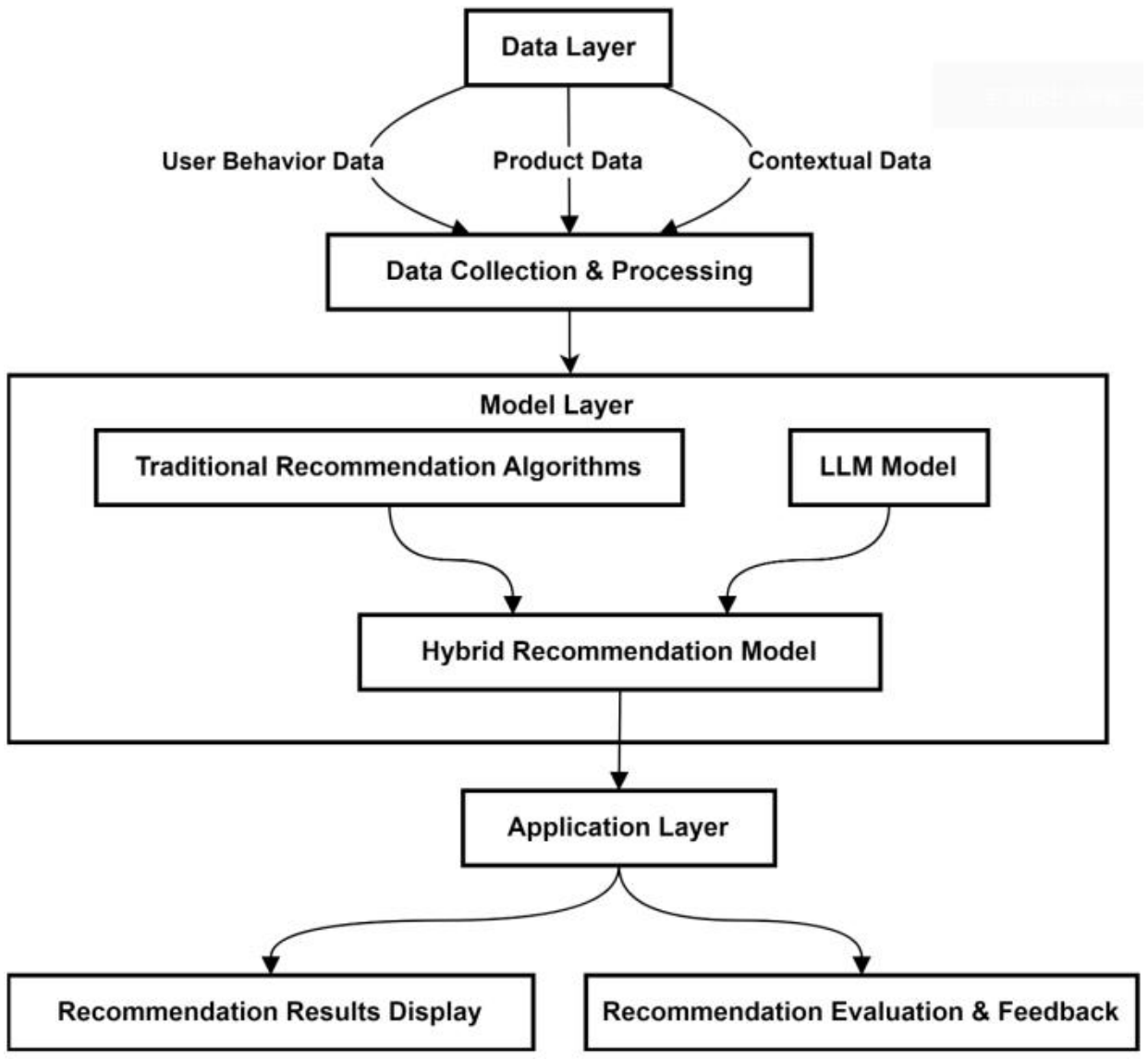

Fig. 2. Research framework design

### B. Data Collection and Processing

This paper collects three types of data: user behavior, product, and context, covering user activity records, product attributes, and environmental information. Data processing includes three steps: cleaning, normalization, and feature extraction. Through these steps, user interest vectors and product feature vectors are extracted from the original data, and the final input features are generated in combination with context information. These processed data provide the basis for subsequent recommendation models.

$$\mathbf{x}_u = \text{ExtractFeatures}(D_u) \tag{2}$$

$$\mathbf{y}_i = \text{ExtractFeatures}(D_i) \tag{3}$$

Among them, $\mathbf{x}_u$ represents the feature vector of user $u$, $\mathbf{y}_i$ represents the feature vector of item $i$, $D_u$ and $D_i$ are the data sets of users and items respectively.

### C. Integrated design of personalized recommendation algorithm and LLM

In the integrated design of personalized recommendation algorithm and LLM, this paper adopts a hybrid model structure, that is, combining LLM with existing collaborative filtering (CF) and content-based recommendation (CBF) algorithms.

#### 1) Content-based recommendation

Generate a preliminary recommendation list by analyzing the text data of users and products. The specific formula is as

$$S_{u,i}^{\text{CBF}} = \cos(\mathbf{x}_u, \mathbf{y}_i) \tag{4}$$

follows:

Among them, $S_{u,i}^{\text{CBF}}$ represents the similarity between user $u$ and product $i$, and $\cos(\mathbf{x}_u, \mathbf{y}_i)$ represents the cosine similarity between the user feature vector and the product feature vector.

#### 2) Collaborative filtering recommendation

Recommendations are made based on the similarity between users. The formula is as follows:

$$S_{u,i}^{\text{CF}} = \sum_{v \in U} \cos(\mathbf{x}_u, \mathbf{x}_v) \cdot R_{v,i} \tag{5}$$

#### 3) LLM recommendation

LLM performs semantic analysis on user behavior data to generate recommendation results. LLM recommendation is calculated using the following formula:

$$S_{u,i}^{\text{LLM}} = \text{LLMScore}(\mathbf{x}_u, \mathbf{y}_i) \tag{6}$$

Among them, $S_{u,i}^{\text{LLM}}$ represents the recommendation score between user $u$ and item $i$ by LLM.

The final recommendation score is calculated from the above three parts:

$$R_{u,i} = \alpha \cdot S_{u,i}^{\text{CBF}} + \beta \cdot S_{u,i}^{\text{CF}} + \gamma \cdot S_{u,i}^{\text{LLM}} \tag{7}$$

## IV. PERSONALIZED RECOMMENDATION MODEL BASED ON LLM

### A. Model architecture design

The architecture and process of the personalized recommendation model based on LLM are shown in Fig. 3:

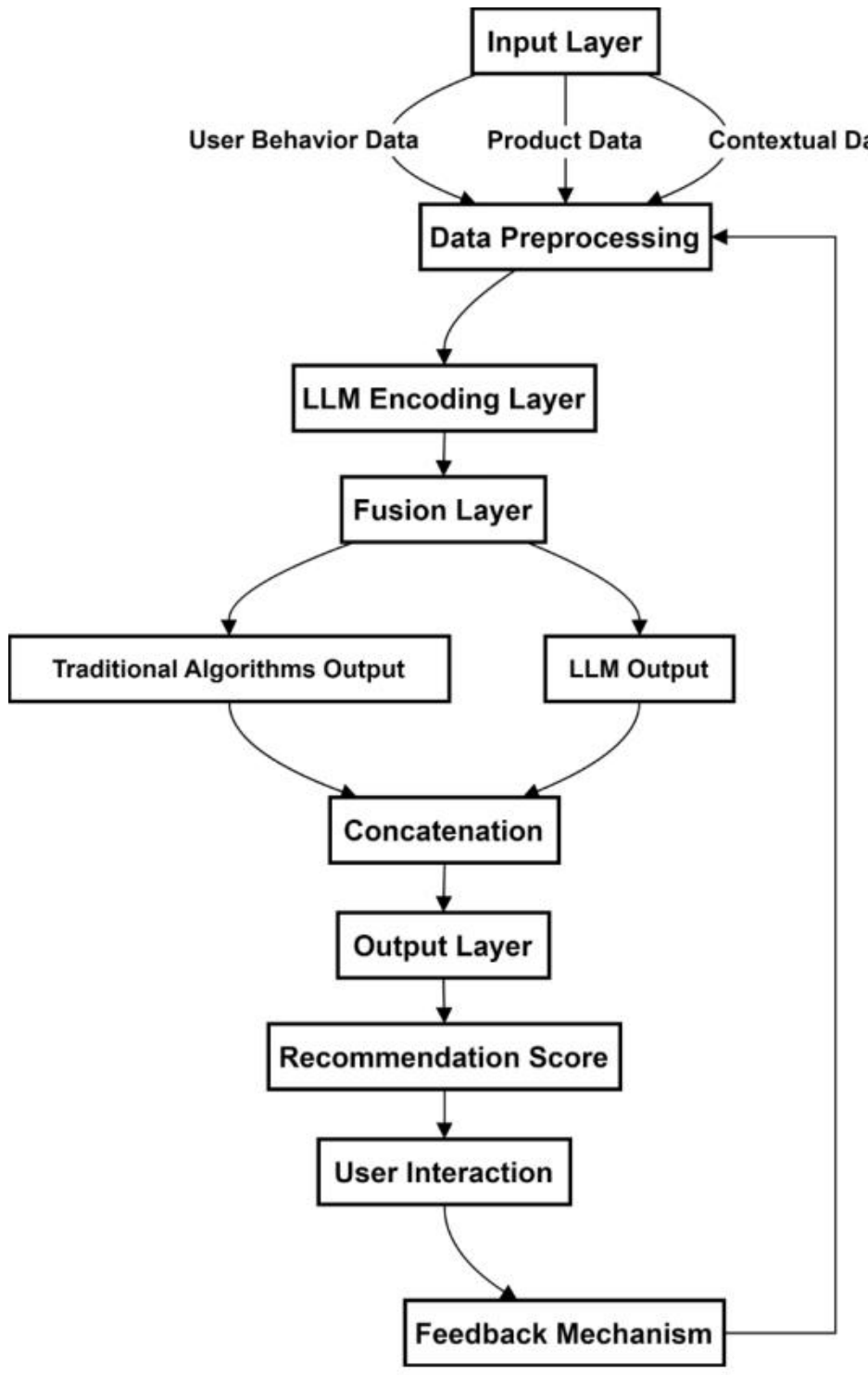

Fig. 3. The process of designing the personalized recommendation model architecture based on LLM

The data flow in the architecture can be expressed by the following formula:

$$R_{u,i} = f_{out}(\text{Concat}(h_{\text{LLM}}(x_u, y_i), h_{\text{trad}}(x_u, y_i))) \quad (8)$$

Where $R_{u,i}$ represents the final recommendation score of user $u$ for product $i$, $h_{\text{LLM}}(x_u, y_i)$ represents the encoding output of LLM for the features of user $u$ and product $i$, $h_{\text{trad}}(x_u, y_i)$ represents the feature output of the traditional recommendation algorithm, Concat represents the feature concatenation operation, and $f_{out}$ represents the scoring function of the output layer.

### B. Model training and optimization

In order to ensure the performance of the model, we adopt the following strategies for model training and optimization:

#### 1) Multi-task learning:

Combine the recommendation task with other related tasks (such as user intent recognition and sentiment analysis) to improve LLM's learning ability for multi-dimensional information by sharing model parameters.

#### 2) Loss function design:

Add diversity constraints based on the traditional mean square error (MSE) or cross entropy loss function to avoid the recommendation results being too single. The loss function can be expressed as:

$$L = L_{\text{main}} + \lambda \cdot L_{\text{diversity}} \quad (9)$$

Where $L_{\text{main}}$ is the loss function of the main task (such as MSE), $L_{\text{diversity}}$ is the diversity loss term, and $\lambda$ is the weight coefficient used to control the influence of the diversity constraint.

## V. EXPERIMENT AND ANALYSIS

### A. Experimental Environment

The configuration of the experimental environment is crucial to the success of model training and testing. Table I lists the hardware and software environment configuration of this experiment in detail.

TABLE I. HARDWARE AND SOFTWARE ENVIRONMENT CONFIGURATION FOR THIS EXPERIMENT

| Component | Specification/Version |
|---|---|
| Operating System | Ubuntu 20.04 LTS |
| CPU | Intel Xeon Gold 6230 |
| GPU | NVIDIA Tesla V100 |
| RAM | 256 GB |
| Storage | 2 TB NVMe SSD |
| Deep Learning Framework | PyTorch 1.9.0 |
| Python Version | Python 3.8 |
| LLM Model | Pretrained BERT (base) |

This experiment used a real dataset from a large e-commerce platform, which contains about 500,000 user behavior records, including clicks, searches, purchases, etc., involving about 50,000 products and their detailed descriptions. The dataset also includes more than 100,000 user comments to enhance the semantic understanding ability of the model.

### B. Experimental Design and Implementation

The experimental design and implementation part follows the framework mentioned above, covering data preprocessing, model training and testing, and integration of user feedback. The experiment uses a real e-commerce dataset to evaluate the performance of the LLM-based personalized recommendation model on multiple indicators.

### C. Experimental Results Analysis

Fig. 4 shows the performance of the traditional model and the LLM-based model across multiple metrics. The color intensity indicates the relative performance, with darker shades representing higher scores. This visualization allows for a quick, comparative analysis of model performance across key evaluation metrics.

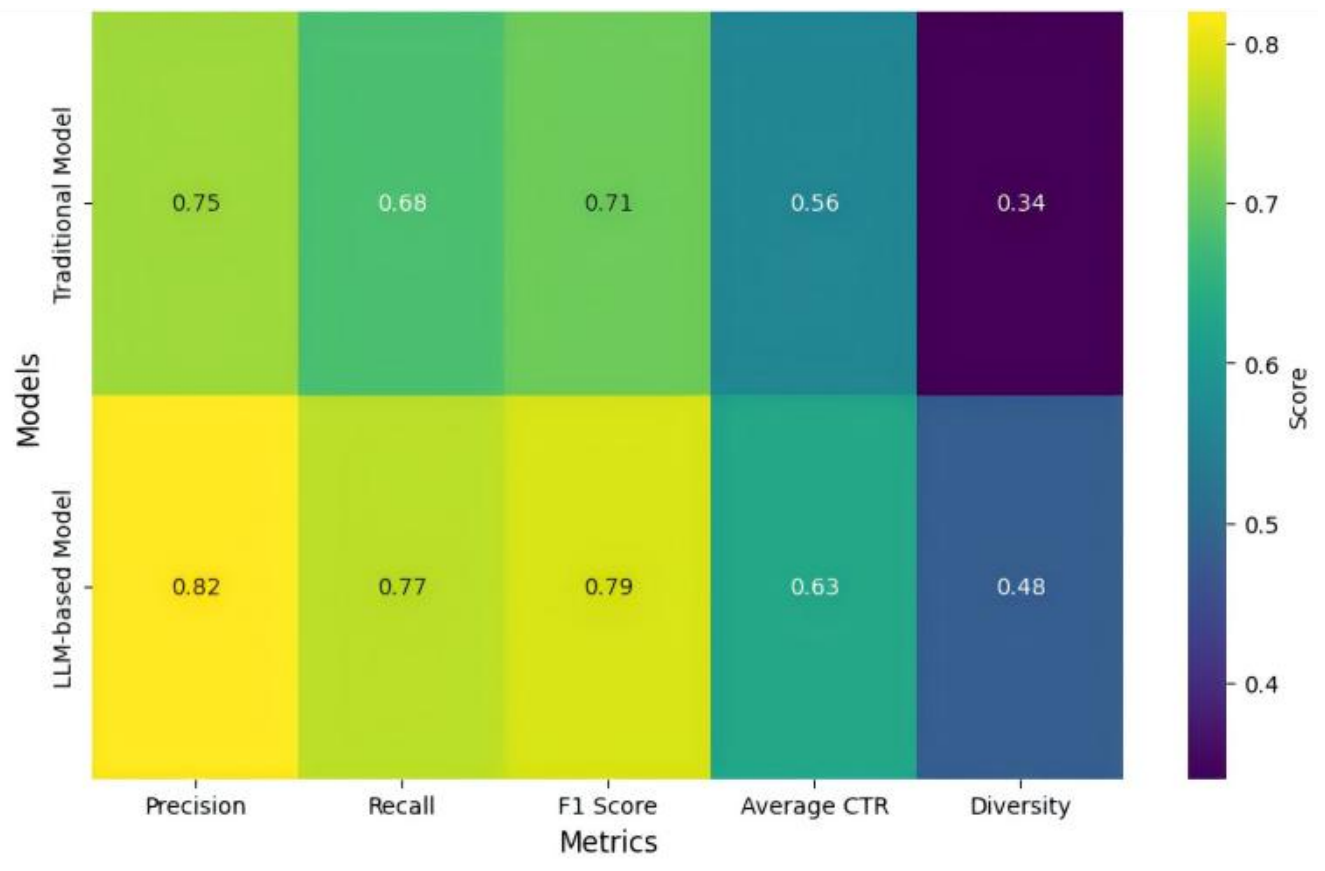

Fig. 4. Multi-Metric Performance Heatmap

The precision-recall curve further illustrates the advantage of the LLM-based model (see Fig. 5), with both precision and recall metrics showing marked improvements over the traditional model, indicating better overall performance in identifying relevant recommendations.

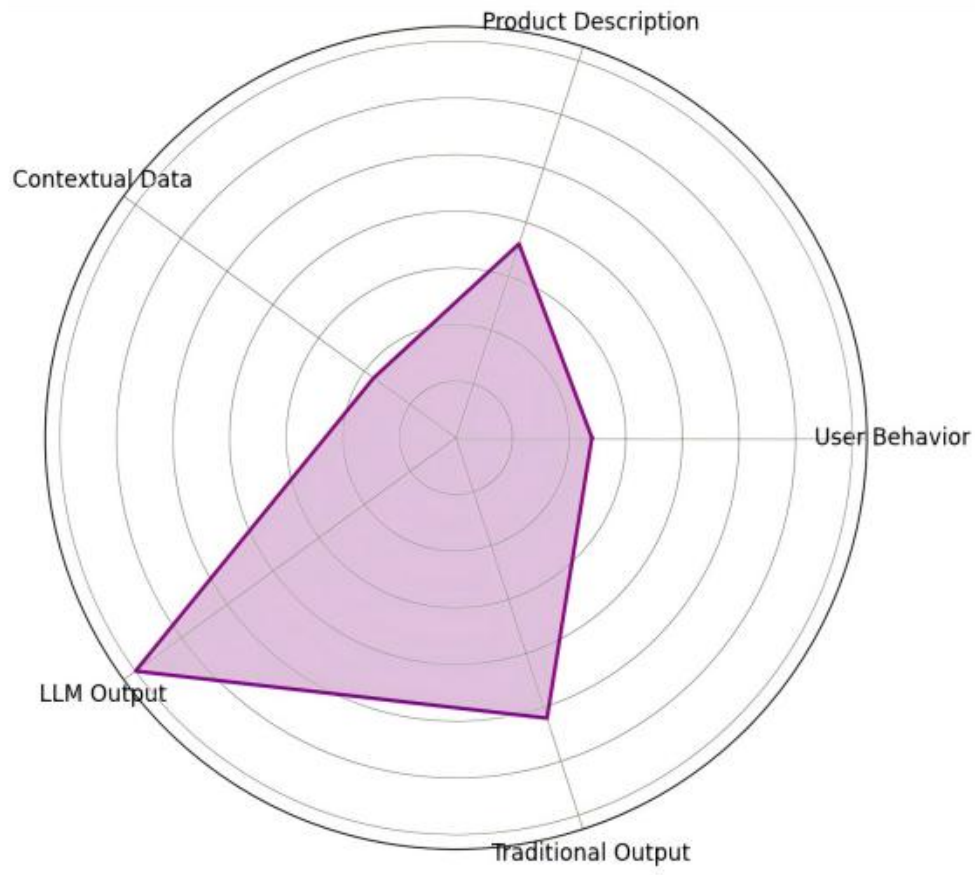

Fig. 5. Precision-Recall Curve for Traditional vs. LLM-based Models

The experimental results show that the personalized recommendation model based on LLM outperforms the traditional recommendation model in many evaluation indicators, especially in terms of accuracy and recommendation diversity. Table II shows the key results of the experiment.

TABLE II. PERFORMANCE COMPARISON BEFORE AND AFTER OPTIMIZATION

| Metric | Traditional Model | LLM-based Model |
|---|---|---|
| Precision | 0.75 | 0.82 |
| Recall | 0.68 | 0.77 |
| F1 Score | 0.71 | 0.79 |
| Average CTR | 0.56 | 0.63 |
| Recommendation Diversity | 0.34 | 0.48 |

## VI. CONCLUSION

This study shows that by introducing a large language model (LLM), we can significantly improve the performance of personalized recommendation systems on e-commerce platforms. The experimental results clearly show that the recommendation model based on LLM outperforms traditional recommendation algorithms in multiple key indicators, such as precision, recall, and recommendation diversity. This improvement is not only due to LLM's deep understanding of natural language, but also its ability to process subtle semantics in user comments and product descriptions, thereby more accurately grasping the real needs of users. In addition, LLM shows high flexibility and robustness in adapting to different user scenarios, and can dynamically adjust the recommendation strategy according to the user's context, making the recommendation results more relevant and personalized.

Overall, this study provides a new idea for personalized recommendation. By combining the advantages of traditional recommendation algorithms with the semantic processing capabilities of LLM, we not only improve the accuracy of recommendations, but also greatly increase the diversity of recommended content. The application of this method can not only improve the user experience, but also significantly improve the sales and user stickiness of e-commerce platforms.

In future work, the integration of deep reinforcement learning (DRL) techniques could be explored to further enhance the adaptability of personalized recommendation systems. DRL's ability to optimize decision-making in dynamic and uncertain environments [22] could complement the semantic understanding capabilities of LLMs. By incorporating DRL, future systems may become more responsive to changing user preferences and behaviors, offering a more dynamic, personalized, and context-aware experience. Additionally, adversarial learning techniques could be investigated to strengthen the robustness of recommendation systems. Zhu et al. [23] demonstrated the effectiveness of adversarial strategies in optimizing sequential recommendations within multi-latent spaces, which could help address challenges related to evolving user interests and data sparsity. By combining DRL and adversarial learning, future systems may better navigate complex user behavior patterns and provide more resilient and adaptive recommendations.